\documentclass[aps,showpacs,amsmath,amssymb,twocolumn,prl,floatfix,superscriptaddress,a4]{revtex4}
\usepackage{graphicx}
\usepackage{epsfig}
\usepackage{amsmath}
\usepackage{amsfonts}
\usepackage{amssymb}

\begin{document}
\title{Fundamental Gates for a Strongly Correlated Two-Electron Quantum Ring}

\author{L. S\ae len}
\affiliation{Department of Physics and Technology, University
 of Bergen, N-5007 Bergen, Norway}

\author{E. Waltersson}
\affiliation{Fysikum, Stockholm University, AlbaNova, S-106
91 Stockholm, Sweden}

\author{J.P. Hansen}
\affiliation{Department of Physics and Technology, University
 of Bergen, N-5007 Bergen, Norway}
 
\author{E. Lindroth}
\affiliation{Fysikum, Stockholm University, AlbaNova, S-106
91 Stockholm, Sweden}

\date{\today}
\pacs{85.35.Be, 03.67.-a ,78.67.-n}


\begin{abstract}
We demonstrate that conditional as well as unconditional basic
operations which are prerequisite for universal quantum gates can be
performed with almost 100\% fidelity within a strongly interacting
two-electron quantum ring. Both sets of operations are based on a
quantum control algorithm that optimizes a driving electromagnetic
pulse for a given quantum gate.  The demonstrated transitions occur on
a time scale much shorter than typical decoherence times of the
system.
\end{abstract}

\maketitle

Quantum computing requires a set  of fundamental single-qubit
operations which can address and manipulate each qubit
regardless of the state of the others. In addition at least one
conditional operation must be defined which can address any chosen
qubit based on the status of another~\cite{Barenco1995}. This poses a
major challenge 
in all logical devices composed of strongly
interacting single particle qubits: The interaction then creates
entangled multi-particle states which hide the single particle
character completely, e.g. regarding the energy spectrum or the
spatial particle distribution. Nevertheless, several basic operations
and quantum information algorithms have been demonstrated,
in nuclear magnetic resonances~\cite{nmr}, trapped
ions~\cite{Cirac1995} and coupled superconducting Josephson
junctions~\cite{super}, but scalability and decoherence remain
severe obstacles in taking experiments from a demonstration level to
manipulation of a large number of qubits.

The original idea to build gates from coupled quantum dots  by
Loss and DiVincenzo~\cite{Loss1998} was based on single electron spin
states interacting in neighboring dots. Recently Petta
\textit{et.~al.}~\cite{Petta2005} demonstrated single qubit control
using the total spin state of a two-electron quantum dot
molecule. Conditional operations in coupled quantum dots have also
been experimentally demonstrated~\cite{Robledo2008}, where excited
states are a part of the information carrier. Another suggestion has
been to include two qubits in a single quantum dot molecule with the
total spin as one qubit and charge localization as the
other~\cite{Hanson2007}. 

Relative to coupled quantum dot-molecules and
quantum dot-arrays, the quantum-ring structure possesses a high-degree of
symmetry,  implying the existence of
conserved quantities, e.g. persistent
currents~\cite{persistent-current},  related to the conservation of
total electron angular momentum. The use of conserved quantities for
the buildup of a quantum processor may be advantageous, compared to
e.g. charge localized states, since the former are
time-independent as long as weak decoherence mechanisms, such as
spin-orbit or hyperfine interactions, can be neglected. In compliance
with this we recently proposed  the two-electron quantum ring total angular momentum and
 total electron spin as a pair of independent qubits~\cite{Waltersson2009}. Since the total
angular momentum is truly multivalued, $M_L = 0,\pm1,\pm2,...$ we
coined this system a ``quMbit''.

In this Letter we show that the total orbital angular momentum
and the total electron spin in the two-electron quantum ring, in spite
of the strong electron-electron interaction, can be coherently and
independently manipulated and that the intended quantum state is obtained with
almost 100\% probability. Hereby successful gate
operations are achieved,  for both the unconditional (NOT) and the
conditional (CNOT) inversion operation. 
An alternative route to scalability can then be
foreseen since the information content of each quantum ring
increases with the number of controllable states. After a short introduction 
we  demonstrate conditional and
unconditional manipulations of the angular momenta and finally
 the unconditional manipulations of the spin are outlined.
 
\begin{figure}[ptb]
\includegraphics[width=\columnwidth]{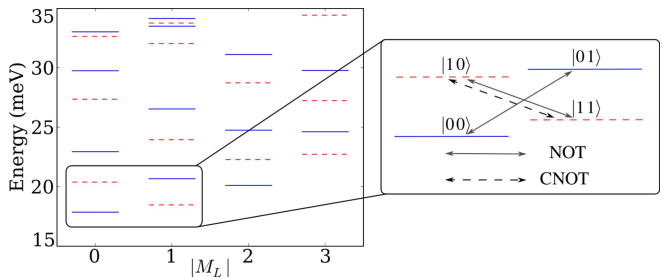}
\caption{The lower part of the energy spectrum of the two-electron quantum ring, cf.~(\ref{fullH0}), as function
of angular momentum, $|M_L|$. Red dashed lines denote triplet states and solid blue lines denote
singlet states. The transition routes for the NOT transitions are highlighted in the right panel. Solid arrows denote the unconditional NOT gate and the dashed arrow denotes the controlled NOT gate. }
\label{fig1}
\end{figure}

The confinement of an electron in a 2D quantum ring
is modeled by a displaced harmonic potential rotated around the
$z$-axis,  giving a two-electron Hamiltonian 
\begin{equation}
\label{fullH0}
	\hat{H}_{0} =  \sum_{i=1,2}  \frac{{\bf p_i}^2}{2m^*}
	+\frac{1}{2}m^*\omega_0^2(r_i-r_0)^2 + 
	\frac{e^{2}}{4\pi\epsilon_{r}\epsilon_{0} r_{12}}.
\end{equation}
Here $m^*$ is the effective mass ($m^* = 0.067m_e$ for GaAs), $\omega_{0}$
determines the confinement strength, $r_i$ is the radial
coordinate for each particle,  $r_{0}$ is the ring radius and $\epsilon_{r}$ is the relative dielectric constant ($\epsilon_{r}=12.4$ for GaAs). We have varied the
ring parameters around the values used in experiments~\cite{Lorke2000}
to optimize the gate performances and settled for $r_{0}=2.5$
a.u.$^{\ast}\approx 24.5$ nm and a potential strength of
$\hbar\omega_0=15$ meV, which have been used throughout this work.
The eigenvectors and 
eigenvalues of the Hamiltonian  (\ref{fullH0}) are found by exact
diagonalization~\cite{Waltersson2007}. Fig.~\ref{fig1} shows the
energy spectrum (left), where red dashed lines denote triplet states,
$S = 1$, and blue solid lines denote singlet states, $S = 0$. The
right panel shows the two pairs of $\mid S |M_L| \rangle$ states that
constitute our gates and the transitions to be controlled:
The NOT gate is a
logical negation operation  which inverts (switches) the
state of the qubit.
Solid gray arrows indicate such a switch of the orbital
angular momentum independently of the spin state. The
controlled NOT (CNOT) gate is indicated 
 by a single black (dashed) arrow. It changes the orbital angular momentum
state for a  spin triplet state, but leaves it unaffected
for a  singlet. Hence, for this operation, the spin state
is the control bit and the angular momentum is the target bit. 
 
To induce the needed transitions between the different eigenstates of
(\ref{fullH0}), the ring is exposed to an 
electromagnetic pulse, an adiabatically varied homogeneous magnetic field in the $z$-direction, $\mathbf{B}_0(t) = (0,0,B_0(t))$,  and a weak inhomogeneous magnetic field, $\mathbf{B}_s(\mathbf{r},t)$, 
\begin{eqnarray}
	\label{eqn:efelt}
	V_{ext}\left({\bf r},t \right) & = & - e{\bf E}\left( t \right)\cdot {\bf r}  + \frac{e^2}{8m^*}B_0^2\left( t \right) r^2\nonumber \\ 
  	& + &   \frac{e}{2m^*} \left( {\bf B}_s \left( {\bf r},t \right) +  \mathbf{B}_0 \left( t \right) \right) \cdot 
 \left( g^*\hat{\mathbf{S}} + \hat{\mathbf{L}} \right)   
\label{eqn:B0}
\end{eqnarray}
resulting in a time dependent Hamiltonian $\hat{H}(t)= \hat{H}_{0} +
V_{ext}({\bf r},t)$. The electric field drives the  angular momentum CNOT- and NOT gate operations, while the two magnetic fields are needed to perform the unconditional spin flip. The inhomogeneous field is typically several orders of magnitude weaker than $B_0$ and is  omitted in the diamagnetic term. 
Through quantum control algorithms~\cite{Rasanen2007, Saelen2008}, the electric pulse $\mathbf{E}$ can be optimized with respect to the desired  gate operation. 
%
%

Numerical solutions of the time dependent Schr\"odinger
equation  were recently used to show that a CNOT operation could be realized
with $\sim 97$~\% fidelity for a $\Delta M_{L}=1$
transition~\cite{Waltersson2009}. For this a circularly polarized  electric field pulse, ${\bf E}(t)=E(t)[\cos{(\omega_{L}t)}
 \hat{\mathrm{{\bf x}}}\pm \sin{(\omega_{L}t)}\hat{\mathrm{ {\bf
    y}}}]$ was  driving the transition between the two
qubit levels, cf.~Fig.~\ref{fig1}.
The transition was realized with a central frequency, $\omega_L$,
corresponding to the energy difference between the active $M_{L}$
states, within a transition time $T=500$ a.u.$^{\ast} \sim28$ ps
and with an intensity $E_{0}\approx0.01$ a.u.$^{\ast}$ $\sim
2.4\cdot10^{2}$W/cm$^{2}$. By optimizing the transition using two
independent electric fields in the $x-$ and $y-$ direction we obtain a
significant improvement in fidelity as well as a shorter 
transition time. The electric field is
defined as a set of piecewise  constant functions on the
divided time interval, $\{E_{t_i}\}$, $t_i \in
[0,t_1,\dots,T_{\rm{final}}]$. During the time 
propagation of the
system wavefunction, $\Psi(t)$, the field components  are adjusted at 
each step  according to a first order
 scheme~\cite{Krotov1996}, e.g. for the x-component
\begin{equation}
	E^{I+1}_{t_i}\hat{\mathbf{x}}  =  - \frac{ \mbox{\rm{Im}}\langle\chi^I(t_i)| e \left(x_1 + x_2\right)|\Psi^{I+1}(t_i)\rangle}{\lambda}\hat{\mathbf{x}},
\label{u(t)}	
\end{equation}
and similarly for the y-component.
In (\ref{u(t)}) $I$ is the iteration number,
and
$\chi^I(t)$ is the solution to the Schr{\"o}dinger equation with
termination condition $\chi^I(T) =
|\Phi_f\rangle\langle\Phi_f|\Psi^I(T)\rangle$. We want to maximize the
projection (the {\em yield}), $\mid \langle\Phi_f|\Psi^I(t)\rangle \mid^2$. The only
constraint in this simple scheme is an energy penalty given by the
parameter $\lambda$ favoring low intensity fields. Additional
penalties on the structure and derivative of the control fields can be
implemented to increase the fidelity even further~\cite{Nepstad2009}.
Here we utilized this possibility for the CNOT.
Through optimization with respect to final states
for {\em both} the singlet and the triplet systems simultaneously, we are in addition
able to achieve very high yields for  the complete CNOT gate as well as for the unconditional
angular momentum flip. The optimization in the former case  is
done  starting in the $M_L=0$ triplet and singlet states, $\Phi_0 = |0\,0\rangle|1\,0\rangle$, and using the combined target state, $\Phi_f = |0\,0\rangle|1\,1\rangle$. 
Similarly for the
unconditional angular momentum flip; we start 
in the singlet and triplet $M_L=0$ states and optimize
with respect to the target state $|0\,1\rangle |1\,1\rangle$.
This requires complete transitions for two separate energy
differences, implying a more complex driving field, with at least
two central  frequencies.

\begin{figure*}[ptb]
\includegraphics[scale=.4]{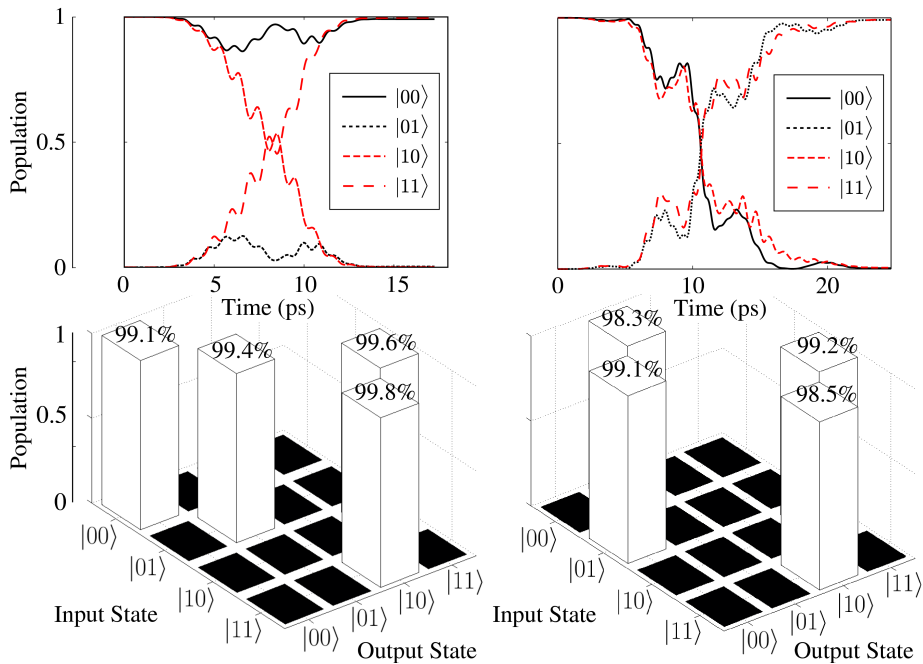}
\caption{ Optimized qubit operations for the angular momentum transition, $M_{L}=0
\leftrightarrow |M_{L}|=1$. Left: 
 two qubit conditional CNOT where the transition
  takes place or not depending on the value of the spin qubit. Right: 
 single qubit NOT, where it always takes place. Depicted  is an initial population
 in $| S|M_{L}|\rangle = |0\,0\rangle$ (solid, black) or $ |1\,0\rangle$ (short dashed,
 red) being transferred to $|0\,1\rangle$ (dotted, black) or
 $|1\,1\rangle$ (long dashed, red). Lower: 
 CNOT (left) and NOT (right) gate truth table at $T_{final}$.
}
\label{fig2}
\end{figure*}
\begin{figure}[ptb]
\includegraphics[scale=.45]{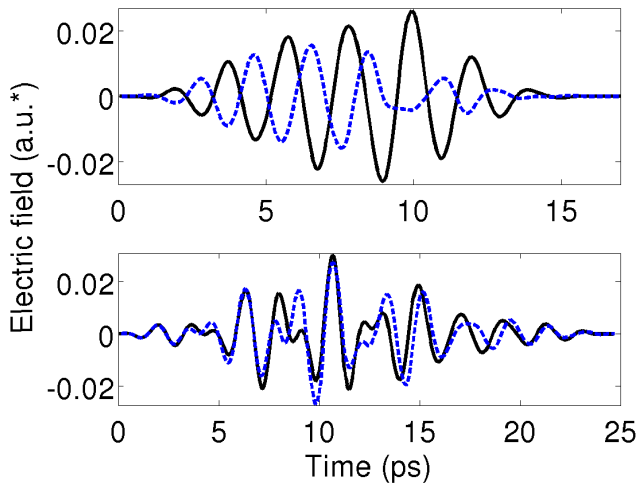}
\caption{Optimized pulses; $x$- (solid, black)
 and $y$-electric field components (dashed, blue) for  CNOT (top) and NOT (bottom)
 gates. }
\label{fig3}
\end{figure}

Fig.~\ref{fig2}  shows the transition dynamics with initial
$M_{L}=0$ population for the two  operations,  the controlled
NOT gate to the left,  and the unconditional NOT gate to the right. 
The upper figures show the population of the qubits during the pulse.
For the CNOT operation the initial state $|S|M_{L}|\rangle =|1\,0\rangle$ is
seen to steadily decay transferring probability to the
$|1\,1\rangle$ state, and eventually we observe a complete transition.
The initial singlet state, $|0\,0\rangle$, on the other hand is
transiently coupled to other states but recovers its initial population
at the end of the pulse. Correspondingly, the upper right panel shows a
 nearly complete transition from  $M_L=0$ to  $|M_L|=1$
 for both spin states. 
An important gate condition is that the
operations should work also in the opposite directions, e.g. for the
CNOT operation the same pulse should transfer initial population in
$|S |M_{L}|\rangle=|1\,1\rangle$ to $|S |M_{L}|\rangle=|1\,0\rangle$ while
leaving the singlet state
population
unaltered. The lower panels of Fig.~\ref{fig2} show the truth tables after
the completed operation for both initial conditions. The transition yields are
seen to be  $99.1\%$  for the NOT gate and $99.8\%$ for the
CNOT gate,  within transition times as short as $25$ ps
and $17$ ps respectively, i.e. three 
orders of magnitude faster than the inverse of the typical
electron-acoustical phonon scattering rates in
GaAs-structures~\cite{Cliemente_scatt_rate}. For the opposite transitions the yields are
slightly lower but all still  $\geq 98.3\%$. For other linear combinations, 
e.g. an initial entangled state $( |00\rangle + |11\rangle )/\sqrt{2}$,
imperfections are found to be less than $4\%$.
These results are achieved within as few as 10 iterations of the type in Eq.~(\ref{u(t)}).
Fig.~\ref{fig3} shows the optimized pulses for the CNOT and NOT
gates decomposed in $x-$ and $y-$components. 
The pulses are rather simple, with  frequency spectra  
centered around the energy of  the resonance transitions.

\begin{figure}[ptb]
\includegraphics[width=\columnwidth]{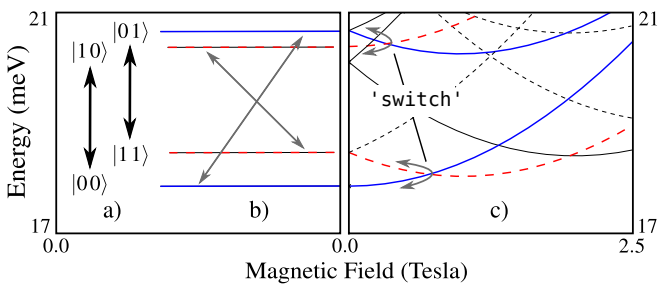}\caption{ $a)$
 Spin flip transitions are indicated by vertical arrows. $b)$ 
  The  qubit states at zero magnetic field. 
 Unconditional angular momentum transitions are indicated by diagonal
 arrows. $c)$ The lower part of the energy
 spectrum  as a function of applied homogeneous
 magnetic field. Solid lines denote singlet states and dashed
 lines triplet states. The qubit states are highlighted with
 thick red and blue curves.}
\label{fig4}
\end{figure}

Finally, we consider unconditional manipulations of the spin. This requires inhomogeneous magnetic fields or other 
spin-mixing interactions.
In principle  an optimization algorithm  as above could be used, simply switching spin with angular momentum. However, present technology cannot realistically deliver inhomogeneous fields stronger than the $mT$ regime across the quantum ring system. We propose therefore a two step procedure involving combined homogeneous and inhomogeneous fields,  illustrated in Fig.~\ref{fig4}. Consider  the transition from  $|S |M_L|\rangle = |0 0\rangle$ to $|1 0\rangle$, i.e. from the lower singlet state (solid blue curve) to the upper triplet state (dashed, red curve).  The right panel shows the energy levels in the presence of a homogeneous magnetic field, obtained by diagonalization of the Hamiltonian~(\ref{fullH0}) in the presence of the $\mathbf{B}_0$ terms in~(\ref{eqn:B0}). The field can bring  the $|00\rangle$-state  adiabatically to the crossing point with the triplet state $|S |M_L|\rangle = |1 1\rangle$. Here the spin is flipped by application of an inhomogeneous magnetic field over the ring. When the homogeneous field is subsequently decreased it is evident that the transition has been made to the 'wrong' triplet state, see Fig.~\ref{fig4}. To make the final transition to the required triplet state $|1 0 \rangle$ we simply apply the unconditional NOT as demonstrated in Fig. \ref{fig2}. The unconditional spin flip operation thus becomes,
\begin{equation}
|\Psi_{S=0/1}(t)\rangle = U_{NOT}^{M_L}(t,t') U_{B_0}(t',0)|\Psi_{S=1/0}(t=0) \rangle
\end{equation}
where $U_{NOT}^{M_L}$ and  $U_{B_0}$ are time evolution operators,
of which  the order is  arbitrary.
If the inhomogeneous magnetic switch is applied at both avoided crossings (curved arrows) the scheme will flip the spin state regardless of initial state. 

The adiabatic development with homogeneous external magnetic fields $B_0$ is well known and the detailed dynamics of the spin flip
transition is now outlined: From decoherence studies it is  known that a weak inhomogeneous magnetic field can flip the spin state of the system. The strength of the magnetic fields is in these cases typically a few mT~\cite{Petta2005,Nepstad2008}. With an inhomogeneous magnetic field  $\sim 10 - 100$~mT, the spin flip can be performed on a much shorter timescale than the natural process. Notably, it has been proposed to selectively flip the spin by making use of the Aharonov-Bohm effect in quantum rings~\cite{Frustaglia2001}. 
 
The two spin states can, with  a circular  configuration,
\begin{equation}
	\mathbf{B}_s(r)\ = \Big\{
		\begin{array}[c]{cc} %
			B_{s}\sin(\phi)\hat{\mathbf{x}} + B_s\cos(\phi)\hat{\mathbf{y}}, & r<r_{0}\\
			0, &  \mbox{otherwise}%
		\end{array}
	 .
\end{equation}

 form a local two-level system.
In the adiabatic basis the dynamics at the avoided crossings is described by,
\begin{equation}
	i \hbar \frac{d}{dt}\left(
	\begin{array}
		[c]{c}
		c_{S=0}\\
		c_{S=1}
	\end{array}
		\right) =\left(
	\begin{array}
		[c]{cc}
		0 & B\\
		B & 0
	\end{array}
		\right) \left(
	\begin{array}
		[c]{c}%
		c_{S=0}\\
		c_{S=1}%
	\end{array}
	\right),
\end{equation}
where $c_S$ denotes the amplitude of the two (avoided) crossing spin
states and $B = \langle S=1 | B_s(r) | S=0 \rangle $ is the coupling
induced by the inhomogeneous field. The spin flip is then realized as
a perfect rotation around the z-axis (on the Bloch sphere) within a
time frame, $\tau\sim\frac{\pi}{2B}$.

 In conclusion we have demonstrated conditional and unconditional fast
 high fidelity quantum gates in a strongly coupled two-electron
 quantum ring model. We remark that the fidelity of each gate may be
 further improved by restricting the upper intensity of the
 controlling fields on the expense of transition times. Alternatively
 it may be increased with fixed intensities and reduced system
 sizes. Storage and control of quantum information has thus been shown
 for two-level spin states entangled with potential multivalued
 angular momentum states.  An extension of the qubit to a multibit may
 be achieved through introduction of higher excitation levels within
 each angular momentum number. The proposal rests on the ability to
 steer the system between initial and final states with close to
 $100\%$ transition probability. This has indeed been achieved with
 relatively simple final pulse shapes.
 
We acknowledge support from the
Norwegian Research Council, the
Swedish Research Council (VR) and from the G{\"o}ran
Gustafsson Foundation.


\end{document}